# Bayesian Compressive Sensing with Circulant Matrix for Spectrum Sensing in Cognitive Radio Networks

Fatima Salahdine[1,2], Naima Kaabouch[1], Hassan El Ghazi[2]
[1]Electrical Engineering Department, University of North Dakota, Grand Forks, USA
[2]STRS Laboratory, National Institute of Posts and Telecommunication, Rabat, Morocco
Email: fatima.salahdine@und.edu, naima.kaabouch@engr.und.edu, elghazi@inpt.ac.ma

*Abstract*—For wideband spectrum sensing, compressive sensing has been proposed as a solution to speed up the high dimensional signals sensing and reduce the computational complexity. Compressive sensing consists of acquiring the essential information from a sparse signal and recovering it at the receiver based on an efficient sampling matrix and a reconstruction technique. In order to deal with the uncertainty, improve the signal acquisition performance, and reduce the randomness during the sensing and reconstruction processes, compressive sensing requires a robust sampling matrix and an efficient reconstruction technique. In this paper, we propose an approach that combines the advantages of a Circulant matrix with Bayesian models. This approach is implemented, extensively tested, and its results have been compared to those of $\mathcal{L}_1$ norm minimization with a Circulant or random matrix based on several metrics. These metrics are Mean Square Error, reconstruction error, correlation, recovery time, sampling time, and processing time. The results show that our technique is faster and more efficient.

*Keywords—Cognitive radio networks; Wideband spectrum sensing; Compressive sensing; Bayesian models; Circulant matrices; $\mathcal{L}_1$ norm minimization.*

## I. INTRODUCTION

Spectrum sensing is one of the main processes of the cognitive radio cycle [1]. In order to sense the wideband radio spectrum, communication systems must use multiple RF frontends simultaneously, which results in long delays, high hardware cost, and computational complexity [2]. To address these problems, fast and efficient sensing techniques are needed. Compressive sensing has been proposed as a low cost solution to speed up the scanning process and reduce the computational complexity. It involves three main processes: sparse representation, encoding, and decoding. During the first process, the signal, $S$, is projected in a sparse basis. During the second process, $S$ is multiplied by a sampling matrix, $M_c$, of $M*N$ elements to extract $M$ samples from $N$ of the signal, $S$, where $M \ll N$. In the last process, the signal is reconstructed from the few $M$ measurements [2-4].

For the encoding process, a number of sampling matrices have been proposed in the literature, including random matrix [5, 6], Circulant matrix [7, 8], Toeplitz matrix [8], and deterministic matrix [9]. Because of their simplicity, more interest has been paid to random matrices. These matrices are randomly generated with independent and identically distributed (i.i.d) elements such as Gaussian and Bernoulli distributions [5, 6]. In general, compressive sensing requires that the sampling matrix satisfies the Restrict Isometry Property (RIP) condition [10]. RIP is a characteristic of orthonormal matrices bounded with a Restrict Isometry Constant (RIC), which is a positive number between 0 and 1 that respects the RIP condition [11]. This condition allows guaranteeing the uniqueness of the reconstructed solution, $\tilde{S}$, during the decoding process. For random matrices, the matrix satisfies the RIP condition for small RIC [5, 6]. However, these matrices require a great deal of processing time and high memory capacity to store the matrix coefficients [7, 8]. Because of the randomness, the results are uncertain, which makes the signal reconstruction inefficient.

Unlike random matrices, Circulant matrices are efficient, fast in terms of signal acquisition, require fewer measurements to perform, and less time to process. [7]. A Circulant matrix is a structured matrix associated and determined using a predefined vector by cyclic permutation [8]. This matrix satisfies the RIP condition for a small number of measurements [12]. Unlike random matrices, Circulant matrices are not universal. Universality means the sampling matrix can be used to compress a signal sparse in any domain. Circulant matrices have been used only with the $\mathcal{L}_1$ norm minimization technique [7, 8, 13].

For the decoding process, a number of algorithms that exploit the sparsity feature of signals have been proposed in the literature [13-23]. A sparse signal can be estimated from a few measurements by solving the undetermined system using three different types of algorithms: Iterative relaxation [13], Greedy [14], and Bayesian models [16]. The iterative relaxation category includes techniques that solve the undetermined system using linear programing. Some techniques classified under this category are $\mathcal{L}_1$ norm minimization, also known as basis pursuit [13], gradient descent [19], and iterative thresholding [20]. Greedy algorithms consist of selecting a local optimal at each step in order to find the global optimum, which corresponds to the estimated signal coefficient. Examples of techniques classified under this category are matching pursuit [21], orthogonal matching pursuit [22], and stage wise orthogonal matching pursuit [23]. Bayesian compressive sensing algorithms consist of using a Bayesian model to estimate the unknown parameters in order to deal with uncertainty in measurements. Examples of techniques classified under this category are Bayesian model using relevance vector machine learning [16], Bayesian model using Laplace priors [17], and Bayesian model via belief propagation [18]. All these Bayesian based algorithms were used only with random matrices.

Iterative relaxation algorithms are more accurate compared to Greedy algorithms, but they are complex, uncertain, require high measurements, and, thus high processing time. Greedy algorithms are fast and require low processing time; however, they are inefficient, uncertain, and require more measurements for the reconstruction process. Bayesian based techniques combine the strengths of both categories. They are fast, accurate, require few measurements for a high recovery rate, and can deal with the uncertainty. In this paper, we propose an approach that combines the strengths of both Circulant matrices and Bayesian models to address the previously mentioned problems during the encoding and decoding processes for fast and efficient compressive sensing. The paper is organized as follows. Section II presents the methodology followed for both encoding and decoding processes as well as the performance evaluation. Section III discusses the simulation results of the proposed approach based on specific metrics. Finally, a conclusion is given at the end.

## II. METHODOLOGY

### A. Bayesian compressive sensing

As previously explained, compressive sensing involves three processes: sparse representation, sensing matrix, and reconstruction, as shown in Fig. 1. In our approach, for the sparsity representation process, we assumed the signal to be sparse. For the sensing matrix process, we used the Circulant matrix. For the reconstruction process, we used the Bayesian model.

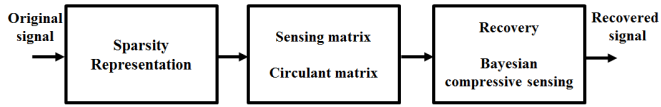

Fig. 1. Block diagram of compressive sensing model.

For the sampling matrix process, a vector, $c$, is given as $(c_0, c_1, ..., c_{N-1})$. The Circulant matrix, $M_c$, is generated from $c$, where $M_{c(i,j)} = M_{c(j-i)} \, mod(N)$ for $i, j = 0...N$. It can be expressed as

$$M_c = \begin{bmatrix} c_0 & c_1 & \cdots & \cdots & c_{N-1} \\ c_{N-1} & c_0 & c_1 & \cdots & c_{N-2} \\ \vdots & \vdots & \vdots & \vdots & \vdots \\ c_1 & c_2 & \cdots & c_{N-1} & c_0 \end{bmatrix} \quad (1)$$

During the encoding process, each column of the matrix is obtained by a right cyclic shift of the preceding column. The values of $c$ are chosen randomly according to a suitable probability distribution to reduce the amount of randomness of the sensing matrix compared to the random matrices. The signal, $S$, is then multiplied by $M_c$ for signal compression. This multiplication is fast because of the reduced number of random coefficients in the Circulant matrix [7, 8].

For the reconstruction process, we used the Bayesian model, which is a probabilistic approach that requires a prior knowledge of parameters to calculate the posterior distribution of the unknown parameters. The Bayesian compressive sensing process consists of finding the sparse solution of a regression problem by exploiting the probabilistic distributions. It solves the undetermined system and finds the accurate solution by estimating efficiently the unknown parameters using the information that we have about the system. It is based on two main elements: the knowledge about the linear relationship between the signal measurements and the original signal, and the knowledge about the fact that the original signal is $k$-sparse. Under the Bayesian model, the $k$-sparse signal is acquired through a product with the Circulant matrix. A noise, $W$, is added to the signal measurements, which includes the noise measurements and the sparse representation error. Fig. 2 illustrates the simulation methodology of our proposed model.

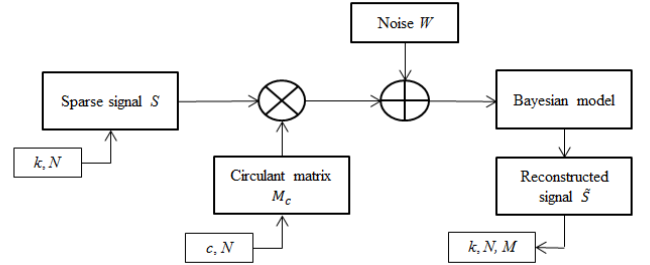

Fig. 2. The model of the Bayesian compressive sensing with Circulant matrix.

The noisy measurements can be formulated as

$$R = M_c \, S + W \quad (2)$$

According to the theorem of central limit for $N \gg M$, $W$ can be approximated as a zero mean Gaussian noise with unknown variance $\delta_w$, which can be expressed as $N(0, \delta_w)$. The signal to be approximated can be considered as a Gaussian variable with $S = (S_1, S_2, ..., S_N)$. Therefore, the Bayesian model implies that the noisy measurements, $R$, is an i.i.d Gaussian and depends on the unknown $S$ and $\delta_w$. This is expressed as

$$R|S, \delta_W \sim N(S, \delta_W) \quad (3)$$

Fig. 3 illustrates the proposed Bayesian model in which the unknown signal, noise, and vector to generate the Circulant sampling matrix are parents of the noisy measurements. Noise variance $\delta_w$, signal mean $\mu_S$, and signal variance $\delta_s$ are the parameters of the noise and the signal that need to be estimated.

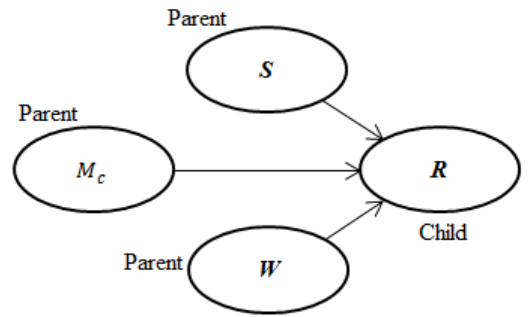

Fig. 3. Graphical model of the Bayesian compressive sensing technique.

The model specifies the conditional probabilities of the measurements $P(R/S)$ and the noise $P(R|\delta_W)$. $P(R/S)$ and $P(R|\delta_W)$ present the probability densities of $R$, as functions of the values taken by the signal and the noise variance respectively. The conditional probability of the signal to be estimated given the measurements can be expressed thought Bayes' rule as

$$P(S/R) = \frac{P(R/S)P(S)}{\sum_{S'} P(R/S')P(S')} \quad (4)$$

where $S'$ represents the other alternative solution of the undetermined system. Consequently, the Gaussian likelihood of the noisy measurements can be expressed from the previous conditional probability as

$$P(R|S, \delta_W) = (2\pi\delta_W)^{-M/2} \exp\left(-\frac{1}{2\delta_W}\right) \|R - M_c S\|^2) \quad (5)$$

Given the Circulant matrix vector $c$, the Circulant sampling matrix coefficients are generated to be used for the Bayesian reconstruction process. This matrix should be the same as the one used for compression to reduce the randomness. With prior knowledge that $S$ is sparse and $M_c$ is known, $S$ and $\delta_w$ are the two quantities to be estimated. The prior density of the unknown signal $P(S/k)$ in terms of sparsity can be expressed as

$$P(S/k) = (k/2)^N e^{(-k \sum_i^N S_i)} \quad (6)$$

where $k$ is a parameter that represents the sparsity of $S = (S_1, S_2,…,S_N)$. Under the Bayesian estimation process, the undetermined system in (2) becomes a linear problem with $S$ sparse and (2) can be reformulated as

$$\tilde{S} = \underset{R = M_c S + W}{\arg\min} \|R - M_c S\|_2^2 + z\|S\|_1 \quad (7)$$

where $z$ is a positive scalar. The objective of our Bayesian model is to look for the posterior probabilistic distribution for $S$ and $\delta_w$ taking into account the known evidences. The maximum posterior probability corresponds to the sparsest solution of the undetermined system presented in equation (7).

The algorithm calculates the joint probability distribution of all unknown parameters and computes the prior distribution of each element of $S$ with the hyper parameters $a$ and $b$. The hyper parameter $a = (a_1, a_2,…, a_N)$ represents the initial posterior of the signal variance and the hyper parameter $b$ represents the initial posterior of the noise variance. The prior distribution of $S$ given the hyper parameters $a$ and $b$ can be expressed as the product of the conjugate prior of signal variance $\Gamma(\delta_{si}/a,b)$ and the likelihood function of $S_i$ which is defined as a zero mean Gaussian prior for each signal coefficients $N(S_i/0, \delta_{wi})$. It is also called the marginal likelihood for Bayes estimation [15]. This probability of the signal given $a$ and $b$ is given as

$$P(S/a,b) = \prod_i^N \int_0^\infty N(S_i/0, \delta_{wi}) \Gamma(\delta_{si}/a,b) d w_i \quad (8)$$

The algorithm optimizes the hyper parameters for the Gaussian process in an iterative loop, estimates new values of $a$ and $b$, and then maximizes the marginal likelihood using the new estimated values of $a$ and $b$. The algorithm is based on the previous results for learning and searching for the new values of the hyper parameters $a$ and $b$. Taking into account the assumption about the knowledge of $a$ and $b$ in addition to $M_c$ and $R$, the posterior probabilistic distribution of $S$ can be then expressed as a Gaussian distribution $S \sim N(\mu_S, \delta_S)$ with mean $\mu_S$ and variance $\delta_S$ which are given by

$$\begin{cases} \mu_S = b\, \delta_S\, M_c^T R \\ \delta_S = (b\, M_c^T M_c + A)^{-1} \end{cases} \quad (9)$$

where $A = \text{diag}(a_1, a_2, …, a_N)$. The last estimated value of $b$ will be the noise variance. At the end of the algorithm, the signal is approximated and the uncertainty is reduced.

### B. Performance evaluation

In order to evaluate the efficiency of our model, the results of this model have been compared to the results of the basis pursuit technique [9] using several metrics. These metrics are: Mean Square Error, reconstruction error, correlation coefficients, processing time, recovery time, and sampling time.

The reconstruction error is a metric that calculates the norm of the difference between the expected signal and the original signal divided by the norm of the original signal. It is expressed as

$$R_e = \frac{\|\tilde{S} - S\|}{\|S\|} \quad (10)$$

Mean Square Error is a metric that measures the average magnitude of the squared difference between the reconstructed signal and the original signal. It corresponds to one of the loss functions used for error estimation. It is expressed as

$$MSE = \frac{1}{N} \sum_N (S - \tilde{S})^2 \quad (11)$$

*MSE* has the same measurement unit as the data being estimated. It is utilized for predictive modeling in order to analyze the variation in the error of the reconstruction algorithm for multiple times.

Correlation measures the similarity between the original signal, $S$, and the reconstructed signal, $\tilde{S}$, to measure how similar they are. The measure of correlation is known as correlation coefficient $C_c$, which is a scalar quantity. It can take values between -1 and 1. It is expressed as

$$C_c = \frac{N \sum(S\tilde{S}) - (\sum S)(\sum \tilde{S})}{\sqrt{N(\sum S^2) - (\sum S)^2} \sqrt{N(\sum \tilde{S}^2) - (\sum \tilde{S})^2}} \quad (12)$$

When $C_c$ is positive and less than 1, it means the two signals are positively correlated and the strength of the correlation is expressed with a percentage value. When $C_c$ is null, it means there is no relationship between the two signals. When $C_c$ is negative and greater than -1, it means the two signals are negatively correlated and the strength of the correlation is expressed with a percentage value.

Recovery time is the time required by the reconstruction process to reconstruct the signal. It allows defining the fastest reconstruction technique. Sampling time is the time required by the sampling matrix process in order to compress the signal using a specific matrix. It allows defining the faster sampling matrix technique. Finally, the processing time is the time required to perform all processes.

## III. RESULTS AND DISCUSSION

The two algorithms, Bayesian compressive sensing and $\mathcal{L}1$ norm minimization with the Circulant matrix, were implemented and extensively tested. Their efficiencies were compared using the metrics previously mentioned ($R_e$, $MSE$, $C_c$, $t_p$, $t_r$ and $t_s$). In this performance evaluation, we investigated the efficiency of the Circulant matrix in sampling signals and compared its results with those of random matrices. We also investigated the performance of our Bayesian model and compared its results with the basis pursuit technique.

Examples of the results are shown in Fig. 4 to 7. Fig. 4(a) shows an example of the original signal with 15 spikes and a total of 200 samples. The noise was added to the original signal and fed to the two algorithms. Fig. 4(b) shows the output signal after applying the Circulant sampling matrix and the Bayesian technique to the signal with added noise. Fig .4(c) represents the output signal after applying random sampling and basis pursuit technique to the original signal with added noise. As one can see in Fig .4(c), the output signal has more fluctuations than the output signal shown in Fig .4(b). These fluctuations correspond to the null coefficients that are non-reconstructed as zero coefficients. Thus, the reconstruction with the Circulant matrix is more efficient compared to the reconstruction with random matrix.

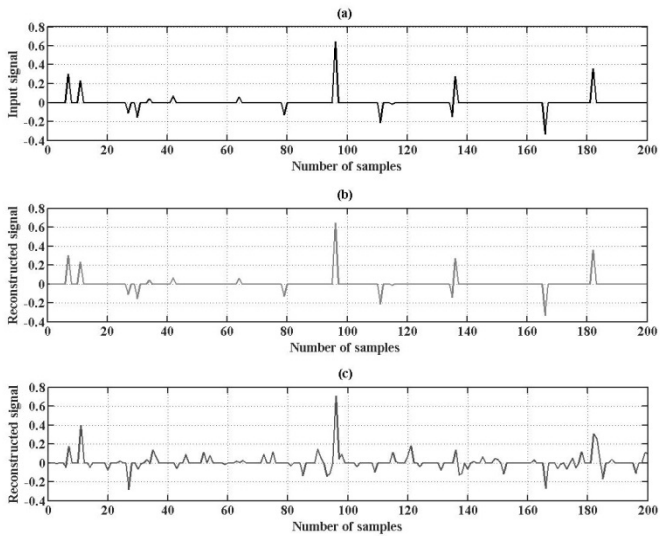

Fig. 4. Example of input signal and the outputs after applying the reconstruction techniques (a) Input signal; (b) Output signal using Bayesian combined with Circulant matrix technique; (c) Output signal using Bayesian combined with random matrix technique.

The sampling time, $t_s$, was computed for each technique. The results show that random sampling matrix requires a great deal of time to process compared to the Circulant matrix. For example, for the signal seen in Fig .4(a), the $t_s$ of Circulant matrix is 0.06 ms while the $t_s$ of random matrix is 0.30 ms. This result shows how dense matrices are slow in terms of computation because of the high required number of measurements and the randomness in their coefficients.

Figs. 5(b) and (c) show the reconstructed signal using basis pursuit and Bayesian model with Circulant matrix technique, respectively. As one can see, for the Bayesian model, the output signal is similar to the original signal and the spikes are completely recovered. However, the output signal of basis pursuit presents more fluctuations than the output signal of the Bayesian model, as shown in Fig. 5(b). Therefore, for high dimensional signal, the Bayesian reconstruction is more efficient than basis pursuit reconstruction. In addition, the sparsity level of the output signal is 14 for the Bayesian technique and 200 for the basis pursuit technique. The basis pursuit technique cannot estimate the exact value of each coefficient of the original signal, and it estimates the zero values as non-zero values with low magnitude. Moreover, the number of measurements to recover the signal needed by each technique is 15 for the Bayesian technique while it is 200 for the basis pursuit technique. Thus, the Bayesian technique is more efficient in reconstructing the original signal and also requires fewer measurements than the basis pursuit algorithm.

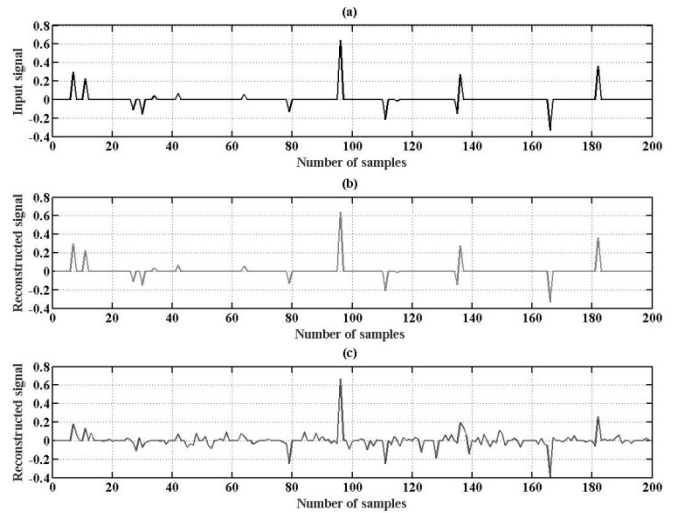

Fig. 5. Example of input signal and the outputs after applying the reconstruction techniques (a) Input signal; (b) Output signal using Bayesian technique combined with Circulant matrix; (c) Output signal using basis pursuit technique combined with Circulant matrix.

Fig. 6 shows the mean square error as a function of the number of samples $N$ for the two reconstruction techniques, Bayesian and basis pursuit. As expected, for both techniques the $MSE$ decreases with the increase of the number of samples. For $N$ from 0 to 100, the Bayesian technique has lower $MSE$ than the basis pursuit algorithm. For higher values of $N$, $MSE$ of both techniques are slightly similar.

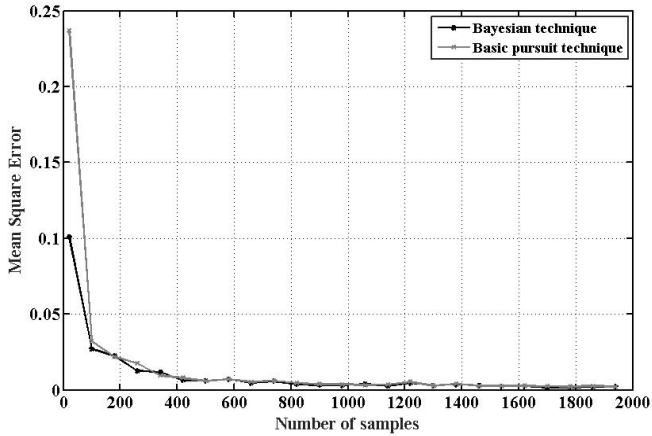

Fig. 6. Mean Square Error as a function of number of samples $N$.

Fig. 7 shows an example of results of the mean square error as a function of the sparsity level for the two techniques, Bayesian and basis pursuit. As can be seen, the *MSE* values corresponding to the Bayesian technique and those corresponding to the basis pursuit technique are slightly similar and increase with the increase of the sparsity. This figure also shows that the more the number of non-zero elements of the signal increases, the more the reconstruction becomes inefficient. One can conclude that the two techniques minimize the *MSE* with the same way with the increase of sparsity level.

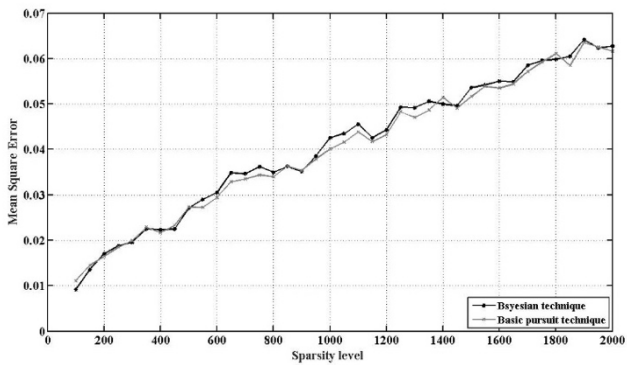

Fig. 7. Mean square reconstruction error as a function of sparsity level $k$.

For the other metrics, Table I gives an example of results of the comparison performance.

As shown in this table, for the reconstruction error $R_e$, the Bayesian technique has an average of 0.77% of reconstruction error. However, the basis pursuit has 6.7% of reconstruction error. Thus, Bayesian is 80 times more precise than basis pursuit with the same matrix. Unlike the basis pursuit, Bayesian technique permits to reconstruct the signal with a very small number errors, which can be explained by the fact that the technique is able to deal with the uncertainty.. Basis pursuit reconstructs the signal with high error level, which can be explained by the fact that this technique cannot handle the uncertainty due to the noisy measurements.

This table also shows that for the correlationmetric, the Bayesian technique presents an average of 100% of correlation while the basis pursuit technique presents an average of 82.87%. Thus, both techniques present a high correlation with values close to 100%, which indicates that the two signals are positively correlated. However, the Bayesian technique presents a better correlation coefficient.

For the recovery time, the Bayesian technique requires an average of 0.90 ms to recover the original signal, but Basis pursuit technique requires 7.20 ms, which represents 12 times higher, thus slower, than the Bayesian technique. For the processing time, Bayesian technique requires an average of 0.96 ms to process while the basis pursuit requires an average of 7. 26 ms. Thus, the Bayesian is 7 times faster than the basis pursuit.

TABLE I. TECHNIQUES COMPARISON BASED ON METRICS

|  | $R_e$ (%) | $C_c$ (%) | $t_r$ (ms) | $t_p$ (ms) |
|---|---|---|---|---|
| **Our technique** | 0.77 | 100.00 | 0.90 | 0.96 |
| **Basis Pursuit with Circulant matrix [6]** | 61.72 | 82.87 | 7.20 | 7.26 |

These examples of results show that the Bayesian technique with Circulant matrix is more accurate, faster, and deals with the uncertainty. In addition, our model requires less measurements, less sampling time, less recovery time, and less processing time. It also allows estimating the original signal with low sparsity level, high correlation, minimizes the Mean Square Error, and handles the uncertainty during the encoding and decoding processes. Thus, our proposed approach includes the strengths of both Bayesian reconstruction and Circulant sampling matrix.

## CONCLUSIONS

In this paper, we have proposed an approach that combines the Circulant sampling matrix with the Bayesian model. This approach allows reducing the randomness and dealing with the uncertainty during the compressive sensing processes. The simulation results have been discussed and compared to those of basis pursuit with Circulant and random matrix techniques. Through comparing and analyzing the simulation results, we can conclude that the Bayesian based algorithm with Circulant matrix is more efficient and fast than the Bayesian with random matrix as well as the basis pursuit with either Circulant or random matrices. For performance evaluation, several metrics have been used, that cover most aspects of the evaluation in terms of reconstruction success, speed, robustness, efficiency, memory, and certainty.